\documentstyle[12pt,openbib]{article}
\begin{document}
\vspace{1cm}
\begin{center}
{\large\bf Glueballs in Peripheral Heavy-Ion Collisions}\\[.4cm]
A.~A.~Natale~\footnotemark\\[.2cm]
\footnotetext{e-mail : natale@axp.ift.unesp.br}
Instituto de F\'{\i}sica Te\'orica,
Universidade Estadual Paulista\\
Rua Pamplona, 145, 01405-900, S\~ao Paulo, SP, Brazil
\end{center}
\thispagestyle{empty}
\vspace{1cm}

\begin{abstract}
We estimate the cross-section for glueball production in peripheral
heavy-ion collisions through two-photon and double-Pomeron exchange, at
energies that will be available at RHIC and LHC. Glueballs will be
produced at large rates, opening the possibility to study decays with
very small branching ratios. In particular, we discuss the
possibility of observing the subprocess $\gamma \gamma (PP)
\rightarrow G \rightarrow \gamma \gamma$.
\end{abstract}
\newpage
The observation of gluon bound states, called gluonia or glueballs, is a
crucial test of quantum chromodynamics. They are predicted in
several theoretical models and expected to be in the range of $1-2$
GeV. These states could be mainly observed in quarkonium decays,
photon-photon collisions, and in diffractive hadron-hadron
scattering through the double-Pomeron exchange~\cite{robson}.

There are some glueball candidates. For example, the $\eta(1440)$ which
was observed in $J/\Psi$ decays, and whose analyses by the
different experimental groups are still contradictory~\cite{eta}.
There is also the case of the $f_2(1720)$
which also generated controversy between the experimental
groups~\cite{edwards}, and, among others, we can also quote
the states $X(1450)$ and $X(1900)$ found recently in
double-Pomeron exchange,
and which have the correct quantum numbers for glueball
candidates~\cite{abatzis}. The status of the glueball
observation will possibly become clear only with the combination
of data from $p-p$ and $e^+e^-$ machines with larger
statistics. Unfortunately, since the glueball width into
two-photons is small, their observation at large rates in
photon-photon collisions will demand new high-luminosity
$e^+e^-$ colliders. In this work we would like to call attention to the
fact that the already planned heavy-ion colliders, RHIC
at Brookhaven~\cite{rhic} and LHC at CERN (operating in the
heavy-ion mode)~\cite{lhc}, may be a powerful source of glueball
production through photon-photon collisions, as well as
double-Pomeron exchange.

It is known that the main interest of relativistic heavy-ion colliders is
in the search of a quark-gluon plasma in central nuclear
reactions. On the other hand peripheral heavy-ion collisions may give rise
to a huge luminosity of photons, opening possibilities of
studying electromagnetic physics, as discussed at length
by Bertulani and Baur~\cite{bbrep}, as well as the
possible discovery of an intermediate-mass Higgs
boson~\cite{papa}, or nonstandard $\gamma\gamma$
processes~\cite{almeida}, which is a physics that requires very energetic
photons. However, it is important to remember that
most of the photons in heavy-ion collisions will
carry only a small fraction of the ion
momentum, favoring low mass final states, and recently we pointed out
that hadronic resonances with $J^{PC}=0^{\pm +},2^{\pm
+},...$ are copiously produced in peripheral
heavy-ion collisions through $\gamma\gamma$ or
double-Pomeron $(PP)$ fusion~\cite{natale}. In
particular, we discussed how a particle as elusive as the $\sigma$
meson, could be observed in the reaction $\gamma\gamma (PP)
\rightarrow \sigma \rightarrow \gamma \gamma$, even
considering the small width of the $\sigma$ into photons.
The same scenario will occur for glueballs, and we will
give here estimates of glueball $(G)$ production in
peripheral collisions at RHIC and LHC, and consider its observation in
the subprocess $\gamma \gamma (PP) \rightarrow G
\rightarrow \gamma \gamma$.

We start our calculation remembering that the photon distribution in the
heavy-ion collision can be determined through
the equivalent photon or Weizs\"{a}cker-Williams
approximation. Denoting by $F(x)dx$ the number of
photons carrying a fraction between $x$ and $x+dx$ of the
total momentum of a nucleus of charge $Ze$, we can define
the two-photon luminosity through
\begin{equation}
\frac{dL}{d\tau}=\int_{\tau}^{1} \frac{dx}{x} \, F(x)F(\tau /x),
\label{e1}
\end{equation}
where $\tau = \hat{s} /s$, and $s(\hat{s})$ is the square of the
c.m.-system (c.m.s) energy of the ion-ion (photon-photon) system. The
total cross section $ZZ \rightarrow ZZ\gamma\gamma \rightarrow ZZG$
can be written as
\begin{equation}
\sigma(s)=\int d\tau \, \frac{dL}{d\tau}\hat{\sigma}(\hat{s}),
\label{e2}
\end{equation}
where $\hat{\sigma}(\hat{s})$ is the cross section of the subprocess
$\gamma\gamma \rightarrow G$.

Taking into account a prescription for photon distribution in
peripheral collisions proposed by Baur~\cite{baur}, we
will use the most realistic photon distribution function determined by
Cahn and Jackson~\cite{papa}, who obtained the following
expression for the differential luminosity:
\begin{equation}
\frac{dL}{d\tau}=\left(\frac{Z^2 \alpha}{\pi}\right)^2 \frac{16}{3\tau}
\xi (z),
\label{e3}
\end{equation}
where $z=2MR\sqrt{\tau}$, $M(R)$ is the nucleus mass(radius), and
$\xi(z)$ is given by
\begin{equation}
\xi(z)=\sum_{i=1}^{3} A_{i} e^{-b_{i}z},
\label{e4}
\end{equation}
which is a fit resulting from the numerical integration of the photon
distribution, accurate to $2\% $ or better for $0.05<z<5.0$, and where
$A_{1}=1.909$, $A_{2}=12.35$, $A_{3}=46.28$, $b_{1}=2.566$,
$b_{2}=4.948$, and $b_{3}=15.21$. For $z<0.05$ we use the expression (see
Cahn and Jackson~\cite{papa})
\begin{equation}
\frac{dL}{d\tau}=\left(\frac{Z^2 \alpha}{\pi}\right)^2
\frac{16}{3\tau}\left(\ln{(\frac{1.234}{z})}\right)^3 .
\label{e5}
\end{equation}
To estimate the glueball production, we note that these states can be
formed by photon-photon fusion with a
coupling strength that is measured by their two-photon
width
\begin{equation}
\hat{\sigma}_{\gamma\gamma \rightarrow G}=(2J+1) \frac{8 \pi^2}{M_G s}
\Gamma_{G \rightarrow \gamma \gamma}
\delta\left(\tau-\frac{M_G^2}{s}\right),
\label{e6}
\end{equation}
where $M_G$ is the glueball mass. There are several
calculations for the
two-photon widths of glueballs~\cite{barnes}, and we will use the
conservative result of Ref.~\cite{kada}. In the case
of the $\eta (1440)$
with $J^P = 0^-$ and $L=S=1$, Ref.~\cite{kada} gives the following result
\begin{equation}
\Gamma (\eta (1440) \rightarrow \gamma \gamma) = \frac{512}{9 \pi^{2}}
{\alpha}^2 {\alpha}_s^2 \frac{1}{{M}_G^4}
{|{R}_1^{\prime}(0)|}^2,
\label{e7}
\end{equation}
where $\alpha ( \alpha_s )$ is the electromagnetic(strong) coupling
constant, and $R_L(r)$ is the radial part of the wave function of the
two-gluon system in configuration space. The calculation goes
through the quark box diagram connecting two-photons to two-gluons
and these will form the bound state. We are stressing this
point because we shall return to it when discussing the double-Pomeron
exchange. The unknown radial wave functions are eliminated
in function of known partial decay widths involving these
gluonium states, arriving at $\Gamma(\eta (1440)
\rightarrow \gamma \gamma)B( \eta(1440) \rightarrow K
\bar{K} \pi)=90 \, eV$, for the $J=0$ and $L=1$ state.
Partial widths for other states can also be found in Ref.~\cite{kada}.

We will compute the production rates only for $\eta (1440)$ and
$f_2(1720)$, and with the average values
of Ref.~\cite{pdg} we obtain $\Gamma (\eta (1440)
\rightarrow \gamma \gamma)= 90 \, eV$, $\Gamma
(f_2(1720) \rightarrow \gamma \gamma)= 223 \, eV$ for the $J=2$
and $L=2$ state, and $13 \, eV$ for the $J=2$ and $L=0$
state. To obtain these values we assumed $B(\eta (1440)
\rightarrow K \bar{K} \pi) \sim 1$ and $B(f_2(1720)
\rightarrow K \bar{K}) \sim 0.38$. It is important to notice
that, for simplicity, we are assuming that
$\eta (1440)$ and $f_2(1720)$ are still good
glueball candidates, and, moreover, they are pure gluonium states,
i.e., without any admixture of $q \bar{q}$ states. Among the many other
candidates for glueballs they have been selected as
typical examples, and the production rates that we will
compute are going to be similar for other states. A
full list of possible candidates, with
their different denominations, can be found in Ref.~\cite{pdg}.

In Table $1$ we show the resulting cross-sections for the states we
discussed above. We considered collisions of ${}^{238}U$ at
RHIC, and ${}^{206}Pb$ at LHC. The
energies involved in these colliders will be $\sqrt{s}=2.0 \times
10^2$ GeV/nucleon at RHIC, and $\sqrt{s}=6.3 \times 10^3$ GeV/nucleon at
LHC, operating with the luminosities ${\cal L}_{RHIC} \simeq 10^{27}
cm^{-2} s^{-1}$ and ${\cal
L}_{LHC} \simeq 10^{28} cm^{-2} s^{-1}$. From Table
$1$ we see, for instance, that we are going to have $4.4
\times 10^5$ and $2.1 \times 10^8$ events/yr of the $f_2(1720)$
state with $L=2$, respectively at RHIC and LHC, assuming
$100\%$ efficiency for the peripheral collision separation and
detection of the final state. The decays of the $f_2(1720)$
into two-photons will barely be observable at RHIC, and we shall have
around $300$ events/yr at LHC. Notice that we are
discussing the two-photon decay because this will be the cleanest
decay to confirm the existence of the glueball, as well as it gives the
possibility to understand the underlying QCD calculation. As far as we
know, we have used the smallest glueball partial widths into
two-photons found in the literature, and we expect
the result of Table $1$ to be a conservative one. The total
number of events is quite large, but, as we shall see, the
situation will improve even more when we consider the strongly
interacting peripheral collision.

We now turn to the case of double-Pomeron glueball production. The
cross-section will be given by the convolution of the Pomeron
distribution function in the nucleus, with the cross-section of
the subprocess $PP \rightarrow G$, i.e.
\begin{equation}
\sigma_{ZZ}^{PP \rightarrow G}= \int \, dx_1 \, \int \, dx_2 \,
F_P(x_1)F_P(x_2) \sigma_{PP}^G(\hat{s}).
\label{e8}
\end{equation}
The Pomeron distribution in a nucleus can be obtained folding the Pomeron
distribution function of a
Pomeron in the nucleon with the elastic nuclear form factor. This has
been worked out in detail by Muller and Schramm~\cite{papa}, and is
given by
\begin{equation}
F_{P}(x)=\left(\frac{3A\beta_0 Q_0^2}{2\pi}\right)^2 \frac{1}{x}
\left(\frac{s'}{m^2}\right)^{2\epsilon} e^{-(\frac{xM}{Q_0})^2},
\label{e9}
\end{equation}
where $A$ is the atomic mass, $\beta_0$ is the quark-pomeron coupling,
and is equal to $\beta_0^2=3.93 GeV^{-2}$. The factor ${s'}^{2\epsilon}$ in
Eq.(9), where $s'$ denotes the invariant mass of the subprocess with
which the pomeron participates,
comes from the Regge behavior of the pomeron, whose
trajectory is given by $\alpha_P (t) = 1 + \epsilon + {\alpha'}_{P} t $,
with $\epsilon=0.085$. In Eq.(9) $Q_0 \approx 60$ MeV determines the
width of the nuclear gaussian form factor used to obtain the Pomeron
distribution.

To compute the subprocess cross-section ($\sigma_{PP}^G$), on the basis
of our present understanding of the QCD structure of the Pomeron is
a very difficult task, and we must make use of models for the
Pomeron in order to do that. We will use the
phenomenological fact that the pomeron couples to quarks like a
isoscalar photon~\cite{land1}. This will allow us to obtain
the Pomeron-Pomeron $\rightarrow$ glueball cross-section from the
photon-photon one. However, it has also been found~\cite{land2} that
it is not always a good approximation to take the
Pomeron-quark-quark vertex to be pointlike. In fact, when either or both
of the two quark legs in this vertex goes far ``off-shell'', the
coupling must decrease. The simplest assumption that agrees with
experiment is to take the Pomeron-quark coupling of the form~\cite{land2}
\begin{equation}
\beta_0(Q^2) = \beta_0 \frac{\mu_0^2}{\mu_0^2 + Q^2},
\label{e10}
\end{equation}
where we will simply assume $Q= \frac{1}{2}
M_G$, and $\mu_0^2 =
1.2~GeV^2$ is a mass  scale characteristic of the pomeron. According to this,
it is easy to verify that we can obtain $\Gamma_{G \rightarrow PP}$ from
$\Gamma_{G \rightarrow \gamma \gamma}$ as long as we substitute
$\alpha^2$ (see, for example, Eq.(\ref{e7})) by
$9\tilde{\beta}_0^4/16\pi^2$, where $\tilde{\beta}_0= \beta_0 (M_G/2)$.

The numerical results of glueball production through double-Pomeron
exchange are presented in Table $2$. The resulting
cross-sections are at least one order of magnitude larger
than the ones originated by photon-photon
collisions. As a comparison, for the $f_2(1720)$ (with
$L=2$) we will have $2.1 \times 10^8$ and $1.9 \times 10^9$
events/year respectively at RHIC and LHC (again, assuming $100\%$
efficiency). Since the branching ratio for two-photon decay of
this state is approximately $1.6 \times 10^{-6}$, we can surely
observe the subprocess $PP \rightarrow G \rightarrow \gamma
\gamma$, as a clear signal for this glueball candidate.

Before discussing a little more about the two-photon signal, let us
digress rapidly about the glueball production through
double-Pomeron exchange. Since we are dealing with not well established
objects in QCD, we can
still make a very simple approximation to estimate $\sigma_{PP}^G$.
We know that the geometrical factorization for
hadronic elastic scattering seems to be a quite good
approximation~\cite{povh}, i. e., the total cross-section for diffractive
scattering of a hadron is
proportional to the radius of this hadron, and we also know that the
strong binding force between two gluons is even larger than the one
between the
triplet of colored quarks, therefore the interaction radius $R_G$ of
glueballs should be determined by the mass $M_{G^0}$ of the
lightest $0^{++}$ state (which we assume to be equal to the
$\eta(1440)$ mass), these two facts together allow us to expect
that the cross-section for $PP \rightarrow G$ can be described by
\begin{equation}
\sigma_{PP}^G \sim \pi R_G^2  \,\,\,\,  , \,\,\,\, R_G \sim
\frac{1}{M_{G^0}}.
\label{e11}
\end{equation}
This may be a rough approximation, although we believe it to be enough to
give an order of magnitude estimate. Notice that the first
calculation is consistent with the Donnachie and
Landshof~\cite{don} model for the Pomeron where it couples preferentially
to quarks, whereas this last one can only be considered as
an upper bound for our estimates. Eq.(\ref{e11}) gives a
total cross-section that depends basically on the energy, for
RHIC we obtain $\sigma_{ZZ}^{PP \rightarrow G}=44 \, mb$ and
for LHC $\sigma_{ZZ}^{PP \rightarrow G}=221 \, mb$,
which are indeed quite large rates. We recall again that
this estimate is to be seen as an upper bound, and the
values of Table $2$ are more appropriate, however,
it is valid in the sense that we still barely know
how is the gluon coupling to the Pomeron, and are
starting to collect more data about the Pomeron distribution in hadrons.

We can now make a few comments about the background to the reaction $ZZ
\rightarrow ZZG \rightarrow ZZ \gamma \gamma$, where
the glueball $G$ is produced predominantly through
double-Pomeron exchange. The main background for this process
will come from the continuum subprocess of photon-photon
scattering (or $PP \rightarrow \gamma \gamma$) through the
box diagram. As discussed in Ref.~\cite{natale} for
the $\sigma$ meson case, we can easily see that
the resonant process is larger than the continuum
one~\footnotemark . There is also the \footnotetext{The box
diagram will be dominated by light quarks, and for
these we can use the asymptotic expression of $\gamma
\gamma$ scattering ($\sigma(s) \sim	20/s$), and integrate it
in a bin centered at the glueball mass, and proportional
to the glueball partial width into two-photons, obtaining a
cross-section smaller than the resonant one with subsequent decay into
two-photons. The $PP \rightarrow \gamma \gamma$
process is computed similarly, as discussed above. Notice that the
interference between the box and resonant diagram is not
important, because on
resonance the two processes are out of phase.} possibility of an
accidental background originating, e.g. from the decay
of the glueball into neutral mesons, and also
possible decay of the last ones into $2\gamma$, where the meson
or the $2\gamma$ are identified as a unique fake
$\gamma$. This is hardly going to happen because, due to
the small glueball mass, the opening angle of the meson
decay will be large enough to be detected with the
calorimeters already in use. Finally, the detectors in
these heavy-ion facilities will be prepared to detect
photons of {\cal O}(1) GeV, because these are also a
signal for the quark-gluon plasma formed in central collisions.

In conclusion, we computed the cross-sections for glueball production in
peripheral heavy-ion collisions. The subprocesses
considered were photon-photon and
double-Pomeron exchange. The rates for glueball
production are very large, and will be dominated by
the Pomeron-Pomeron scattering. Our estimates were
conservative, in the sense that we used
the smallest partial widths of glueballs into
two-photons. We also used the Donnachie-Landshoff model
for the double-Pomeron calculation. A naive
calculation using a geometrical model for Pomeron
elastic scattering gives much larger rates. We have not
considered mixture of the glueball with $q\bar{q}$ states, and if there is a
mixing we can expect an increase of the total cross-section.
We call attention to the fact that the
subprocess $PP \rightarrow G \rightarrow \gamma \gamma$ can be
observed, with more than $300$ events/yr already at
RHIC assuming $100\%$ efficiency, and will
provide a very clean signal for the glueball, as
well as the possibility for studying the underlying QCD process.

\section*{Acknowledgments}
I have benefited from discussions with C.~O.~Escobar, R.~Z.~Funchal and
F.~Halzen. The author also acknowledges the hospitality at the International
Centre for Theoretical Physics where part of this work has been done. This
research was supported in part by
the Conselho Nacional de Desenvolvimento Cientifico e Tecnologico (CNPq).
\newpage
\section*{Tables}
\vskip 2.5cm

\begin{table}[hb]
\center
\begin{tabular}{||c|c|c||}
\hline
$Glueball     $          &$\sigma_{RHIC} (mb)$&$\sigma_{LHC} (mb)$\\
\hline
$\eta(1440)$\,$[J=0,L=1]$&$3.0 \times 10^{-3}$&$0.10$  \\
$f_2(1720)$\,$[J=2,L=2]$ &$1.4 \times 10^{-2}$&$0.65$ \\
$f_2(1720)$\,$[J=2,L=0]$ &$8.3 \times 10^{-4}$&$0.38 \times 10^{-1}$ \\
\hline
\end{tabular}
\label{table1}
\caption{Cross-sections for glueball production through photon-photon
fusion. The values are in mb.}
\end{table}
\vskip 2.5cm

\begin{table}[hb]
\center
\begin{tabular}{||c|c|c||}
\hline
$Glueball$               &$\sigma_{RHIC}(mb)$&$\sigma_{LHC} (mb)$ \\
\hline
$\eta(1440)$\,$[J=0,L=1]$&   $0.27$          &$ 1.3$  \\
$f_2(1720)$ \,$[J=2,L=2]$& $1.2$             &$ 6.1$  \\
$f_2(1720)$ \,$[J=2,L=0]$& $0.07$            &$ 0.36$ \\
\hline
\end{tabular}
\label{table2}
\caption{Cross-sections for glueball production through double-Pomeron
exchange. The values are in mb. The nuclei and energies
are the same as in Table $1$.}
\end{table}

\end{document}